\documentstyle[preprint,tighten,aps,epsf,floats]{revtex} 

\def\lqcd{\Lambda_{\rm QCD}}
\def\MSbar{\overline{\rm MS}}
\def\psla{p\!\!\!\slash}
\def\qsla{q\!\!\!\slash}

\begin{document}

\preprint{\tighten \vbox{\hbox{UCSD/PTH 98-32} \hbox{CERN-TH/98-334} 
  \hbox{hep-ph/9811239} \hbox{} }}

\title{$B$ decays in the upsilon expansion}

\author{Andre H.\ Hoang,${}^a$ Zoltan Ligeti,${}^{b,c}$
  and Aneesh V.\ Manohar${}^c$ }
\address{ \vbox{\vskip 0.truecm}
  $^a$Theory Division, CERN, CH-1211 Geneva 23, Switzerland \\[6pt]
  $^b$Theory Group, Fermilab, P.O. Box 500, Batavia, IL 60510\\[6pt]
  $^c$Department of Physics, University of California at San Diego,\\
    9500 Gilman Drive, La Jolla, CA 92093--0319 \\[6pt] }
\date{November 1998}
\maketitle

\begin{abstract}%
Theoretical predictions for $B$ decay rates are rewritten in terms of the
$\Upsilon(1S)$ meson mass instead of the $b$ quark mass, using a modified
perturbation expansion.  The theoretical consistency of this expansion is shown
both at low and high orders.  Our method improves the behavior of the
perturbation series for semileptonic and nonleptonic inclusive decay modes, as
well as for exclusive decay form factors.  The results are applied to the
determination of the semileptonic $B$ branching ratio, charm counting, the
ratio of $B\to X\tau\bar\nu$ and $B\to Xe\bar\nu$ decay rates, and form factor
ratios in $B\to D^{*} e\bar\nu$ decay. We also comment on why it is not
possible to separate perturbative and nonperturbative effects in QCD.
\end{abstract}

\newpage

\section{Introduction}

Heavy hadron decays can be computed using a systematic expansion in powers of
$\alpha_s(m_Q)$ and $\lqcd/m_Q$, where $m_Q$ is the mass of the heavy quark and
$\lqcd$ is the nonperturbative scale parameter of the strong interactions. In
the $m_Q\to\infty$ limit, inclusive decay rates are given by free quark decay
and the order $\lqcd/m_Q$ corrections vanish~\cite{CGG,incl}. The leading
nonperturbative corrections of order $\lqcd^2/m_Q^2$ are parameterized by two
hadronic matrix elements~\cite{incl,BSUV,MaWi,inclnl,Blok} $\lambda_1$ and
$\lambda_2$. These results are now used to determine the CKM matrix elements
$|V_{cb}|$ and $|V_{ub}|$, using experimental data on inclusive semileptonic
$B$ meson decays. For exclusive decays at lowest order in $1/m_Q$, one can
compute form factors in terms of a perturbative series in $\alpha_s$ times the
Isgur-Wise function~\cite{HQS}. Nonperturbative corrections are parameterized
by hadronic matrix elements, and are of order $(\lqcd/m_Q)^n$. In some cases at
zero recoil, the order $1/m_Q$ correction vanishes~\cite{luke}.

At present, the largest theoretical uncertainties in the computation of
inclusive $B\to X_c e\bar\nu$ and $B\to X_u e\bar\nu$ decay rates arise from
poor knowledge of quark masses. The pole mass of a heavy quark is an infrared
(IR) sensitive quantity, and is not well defined beyond perturbation
theory~\cite{ren1,ren2}. This is related to the bad behavior of perturbative
corrections to the inclusive decay rate when it is written in terms of the pole
mass~\cite{LSW,BBB}. A simple method of avoiding problems with the quark mass
is to use instead the $B$ mesons mass. Unfortunately, the $B$ meson and $b$ quark
masses differ by order $\lqcd$, and so this reintroduces a $\lqcd/m_b$
correction to the inclusive decay rate. The decay rate has also been rewritten
in terms of the infrared safe $\MSbar$ mass of the $b$ quark, but the
uncertainties still remain sizable because the value of the $\MSbar$ mass is
not known very accurately, and because the first few terms in the perturbation
series still remain large. 

In this paper, the theoretical predictions for semileptonic $B$ decays are
rewritten in terms of the $\Upsilon(1S)$ meson mass rather than the $b$ quark
mass. At the same time a modified perturbation expansion (referred to as the
upsilon expansion) must be used, which is discussed in detail in Sec.~II. This
procedure eliminates the uncertainty due to the $m_b^5$ factor in the decay
rates, and at the same time improves the behavior of the perturbation series. 
Our formulae relate measurable quantities to one another and the resulting
perturbation series is free of renormalon ambiguities. We will also see
numerically that the   perturbative corrections are small when the $B$ decay
rate is written in terms of the $\Upsilon$ mass. We have discussed the
procedure  in an earlier publication~\cite{prl}, where it was applied to
inclusive semileptonic $B$ decays. In this article, we describe the method in
more detail and also explain why it leads to an improvement in the perturbation
series expansion. The method is applied to inclusive nonleptonic decays as well
as to exclusive decays. The perturbation series is better behaved for all the
processes we have examined. The semileptonic and nonleptonic decay computations
can be combined to obtain new values for the semileptonic branching ratio and
the average charm multiplicity in $B$ meson decay.

The outline of this paper is as follows. In Section II we define the upsilon
expansion.  In Sec.~III we show that it is the theoretically consistent way of
eliminating the $b$ quark mass from the combination of the perturbation theory
results for $B$ decay rates and the $\Upsilon$ mass.   We also comment on the
role of nonperturbative effects in the theoretical expression for the
$\Upsilon$ mass. Exclusive decays are discussed in~IV. In Sec.~V we give the
results for inclusive semileptonic and nonleptonic decay rates in the upsilon
expansion, the ratio of decay rates $R_\tau = {\cal B}(B\to X_c\tau\bar\nu) /
{\cal B}(B\to X_c e\bar\nu)$, the semileptonic $B$ branching ratio,  and charm
counting in $B$ decay. Sec.~VI contains our conclusions and comments on further
applications of our results.

\section[]{The upsilon expansion\footnote{$B$ \lowercase{decays written in
terms of the $\Upsilon$ mass should be referred to as
upsilon-expanded, even though the expansion parameter will be denoted
by $\epsilon$. 
\uppercase{B}lame this on \uppercase{D}onald \uppercase{K}nuth (or on the
\uppercase{G}reeks?), as $\upsilon$~({\tt \${}$\backslash$upsilon\$}) is
similar to $v$~({\tt \$v\$}) in \TeX.}}}

The results for the OPE calculation of inclusive $B$ decay rates can be 
schematically written as
\begin{eqnarray}\label{schemrate}
\Gamma(B\to X_{ijk}) = {G_F^2 \left|V_{\rm CKM}\right|^2\over 192\pi^3}\, 
m_b^5\, (PS)\,
  \bigg[ 1 &+& {\alpha_s\over\pi}\, \epsilon 
  + {\alpha_s^2\over\pi^2} \Big( \beta_0 + 1 \Big) \epsilon^2 
  + {\alpha_s^3\over\pi^3} \Big( \beta_0^2 + \ldots \Big) \epsilon^3
  + \ldots \nonumber\\*
&+& {1\over m_b^2}\, \Big(\lambda_1 + \lambda_2 \Big) + \ldots \bigg] ,
\end{eqnarray}
where the precise coefficients are not shown.  ${ijk}$ denotes an arbitrary
final state created by the $b\to ijk$ transition at short distance (e.g.,
$ijk=c\ell\bar\nu,\ c\bar cs,$ etc.), $V_{\rm CKM}$ contains CKM matrix
elements, $PS$ is the $b$ quark decay phase space at tree level including color
factors (e.g., $ijk=u\bar ud$ is a factor 3 enhanced compared to
$ijk=ue\bar\nu$), $m_b$ is the $b$ quark pole mass,  $\beta_0=11-2n_f/3$ is the
first coefficient of the QCD $\beta$-function, and $\alpha_s$ is the running
coupling constant in the $\MSbar$ scheme at the scale $\mu$. It is convenient
to expand the coefficient of a given $\alpha_s$ order in powers of $\beta_0$ 
which is related to the Brodsky--Lepage--Mackenzie (BLM)
prescription~\cite{BLM}.\footnote{ The subscript BLM will be used to denote
terms of order $\epsilon^n \beta_0^{n-1}$ in the perturbation expansion. In
some cases, the entire $\epsilon^2$ contribution has not been computed, and
only the BLM piece is known. In cases where the entire $\epsilon^2$ term is
known, the BLM contribution is about ten times larger than the non-BLM part.}
For nonleptonic and rare $B$ decays there is another large parameter in
Eq.~(\ref{schemrate}), $\ln(m_W/m_b)\sim2.8$, which will be discussed in
Sec.~IV. These large logarithms can be summed using the renormalization group
equations. The reason for introducing the variable $\epsilon=1$ is explained in
the next paragraph.

Schematically, the perturbative expansion of the $\Upsilon$ mass in terms of 
$m_b$ is
\begin{eqnarray}\label{schemups}
{m_\Upsilon \over 2m_b} \sim 1 - {(\alpha_s C_F)^2\over8}\, \bigg\{ 1 \epsilon
  &+& {\alpha_s\over\pi}\, \Big[ \Big( \ell + 1 \Big) \beta_0 + 1 \Big] 
  \epsilon^2 \nonumber\\*
&+& {\alpha_s^2\over\pi^2}\,
  \Big[ \Big( \ell^2 + \ell + 1 \Big) \beta_0^2 + \ldots \Big] \epsilon^3 
  + \ldots \bigg\} \,,
\end{eqnarray}
where $\ell=\ln[\mu/(m_b\alpha_s C_F)]$, $C_F=4/3$, and the precise
coefficients are again not shown. Note that this series is of the
form $\{\alpha_s^2,\ \alpha_s^3\beta_0,\ \alpha_s^4\beta_0^2,\ \ldots\}$,
whereas the corrections in Eq.~(\ref{schemrate}) are of order $\{\alpha_s,\
\alpha_s^2\beta_0,\ \alpha_s^3\beta_0^2,\ \ldots\}$.  We will show in Sec.~III
that to combine these two equations in a theoretically consistent manner, terms
of order $\alpha_s^n$ in Eq.~(\ref{schemups}) should be viewed as if they were
only of order $\alpha_s^{n-1}$.  For this reason, the power of $\epsilon$ in
Eq.~(\ref{schemups}) is one less than the power of $\alpha_s$.  It is also
convenient to choose the same renormalization scale, $\mu$, when
Eqs.~(\ref{schemrate}) and (\ref{schemups}) are combined. The prescription of
counting $[\alpha_s(m_b)]^n$ in the $B$ decay rate as order $\epsilon^n$, and
$[\alpha_s(m_b)]^n$ in the $\Upsilon$ mass as order $\epsilon^{n-1}$ will be
called the upsilon expansion. It is expected that the infrared sensitivity
present separately in Eqs.~(\ref{schemrate}) and (\ref{schemups}) will cancel
to all orders in perturbation theory in $\epsilon$. Some arguments that support
this will be given in the next section. In the upsilon expansion 
Eq.~(\ref{schemrate}) takes the form
\begin{equation}\label{expdef}
\Gamma(B\to X_{ijk}) = {G_F^2 \left| V_{\rm CKM}\right |^2 \over 192\pi^3}\, 
  \bigg({m_\Upsilon\over2}\bigg)^5\, (PS')\,
  \bigg[ 1 + \epsilon + \Big( \beta_0 + 1 \Big) \epsilon^2 + \ldots \bigg] ,
\end{equation}
where $PS'$ means that the $b$ decay phase space is also expanded in
$\epsilon$.  We emphasize that the coefficients in the series in $\epsilon$
contain different orders in $\alpha_s$. We show in Sec. III that this is
theoretically consistent. Our results in Sec.~IV and V demonstrate that this
expansion simultaneously eliminates the uncertainty in the theoretical
predictions related to the dependence on the poorly known quark masses and also
improves the behavior of the perturbation series.

The precise form of Eq.~(\ref{schemrate}) depends on the decay channel under
consideration. The expression for the $\Upsilon$ mass in terms of $m_b$  is
given by~\cite{PY,Upsmass},
\begin{eqnarray}\label{upsmass}
{m_\Upsilon \over 2m_b} = 1 - {(\alpha_s C_F)^2\over8} \bigg\{ 1 \epsilon 
&+& {\alpha_s\over\pi} \bigg[\bigg( \ell + \frac{11}6\bigg) \beta_0 
  - 4 \bigg] \epsilon^2  \\
&+& \bigg({\alpha_s\beta_0\over2\pi}\bigg)^2 
  \bigg( 3\ell^2 +9\ell +2\zeta(3)+\frac{\pi^2}6+\frac{77}{12}\bigg) 
  \epsilon^3 +\ldots \bigg\} . \nonumber
\end{eqnarray}
The ellipsis denote terms of order $\alpha_s^5$, and non-BLM terms of order
$\alpha_s^4$. For $\mu$ of order $m_b$, Eq.~(\ref{upsmass}) shows no sign of
convergence; for $\mu=m_b$ it yields $m_\Upsilon = 2m_b (1 - 0.011\epsilon -
0.016\epsilon^2 - 0.024_{\rm BLM}\epsilon^3 - \ldots)$, using
$\alpha_s(m_b)=0.22$. (This value of $\alpha_s(m_b)$ will be used whenever we
present numerical results.) The bad behavior of this series is unimportant (it
would be somewhat better in terms of the $\MSbar$ mass at a low scale).  The
only physical question is what happens when we use Eq.~(\ref{upsmass}) to
predict $B$ decay rates in terms of $m_\Upsilon$.

Finally, decays with a charm quark in the final state depend both on $m_b$ and
$m_c$.  It is convenient to express these decay rates in terms of $m_\Upsilon$
and $\lambda_1$ instead of $m_b$ and $m_c$, using Eq.~(\ref{upsmass}) and
\begin{equation}\label{mbmc}
m_b - m_c = \overline{m}_B - \overline{m}_D + \bigg( 
  {\lambda_1\over 2\overline{m}_B} - {\lambda_1\over 2\overline{m}_D} \bigg) 
  + \ldots \,,
\end{equation}
where $\overline{m}_B = (3m_{B^*}+m_B)/4=5.313\,$GeV and $\overline{m}_D =
(3m_{D^*}+m_D)/4=1.973\,$GeV.   One could imagine replacing $m_c$ by the
$J/\psi$ mass using a procedure similar to that for the $b$ quark. However, the
nonperturbative errors in this case are not under control.

\section{Theoretical consistency}

In Ref.~\cite{prl} we showed that our method of combining the expansion of
inclusive $B$ decay rates with the expansion of the $\Upsilon$ mass to
eliminate the dependence on the $b$ quark mass is consistent at large orders in
perturbation theory in the BLM approximation.  In this section we briefly
repeat this argument (Sec.\ A) and show that the upsilon expansion is also
consistent at first and second order in $\epsilon$, including non-Abelian
contributions (Sec.\ B). Our arguments prove that, as far as the cancellation
of infrared sensitive contributions in the perturbative expansion of the $B$
decay is concerned, the upsilon expansion is equivalent to the use of a
short-distance mass like the  $\MSbar$ mass. They also explain why the upsilon
expansion requires powers of $\alpha_s$ to be counted differently in
$m_\Upsilon$ and $B$ decay rates.
We also comment on the relation
between perturbative and nonperturbative effects in the $\Upsilon$ mass (Sec.\
C).

\subsection{Renormalons in the large $\beta_0$ approximation}

The pole mass $m_b$ is an infrared sensitive quantity, which is not well
defined beyond perturbation theory.  It can be related to an infrared safe
mass, such as the $\MSbar$ mass $\overline{m}_b$ via (for $n_f=4$)~\cite{mass}
\begin{equation}\label{polemass}
{m_b \over \overline{m}_b(m_b)} = 1 + {4\alpha_s\over3\pi}\, \epsilon
  + (1.56\beta_0 - 1.07) {\alpha_s^2\over\pi^2}\, \epsilon^2 + \ldots 
\end{equation}
This relation has terms of the form $\alpha_s^n\beta_0^{n-1}n!$ at high orders,
leading to a renormalon ambiguity.  The Borel transform of the  perturbation
series relating the pole mass to the $\MSbar$ mass is~\cite{ren1,ren2}
\begin{equation}\label{borelmass}
\widetilde m_b(u) = \overline{m}_b \bigg\{ \delta(u) +
  {C_F\over\beta_0} \bigg[6e^{-Cu} \bigg({\mu\over m_b}\bigg)^{2u} (1-u)\,
  {\Gamma(u)\,\Gamma(1-2u)\over\Gamma(3-u)} - \frac3u + \ldots \bigg] \bigg\} ,
\end{equation}
where $C$ is a scheme dependent constant ($C=-5/3$ in $\MSbar$). The leading
renormalon ambiguity is given by the $u=1/2$ pole, yielding
\begin{equation}\label{delmass}
  \Delta m_b = - {2C_F\over\beta_0}\, e^{-C/2} \lqcd \,.
\end{equation}

The $\alpha_s$ perturbation series for the inclusive decay rate written in
terms of the pole mass also has a renormalon singularity. As a result, the
inclusive decay perturbation series is badly behaved at high orders.
Renormalons cancel when the inclusive decay rate is written in terms of the
$\MSbar$ mass~\cite{rencan1,rencan2,rencan3,SAZ}, rather than the pole mass.
While this cancellation is present at high orders, the perturbation series
written in terms of the $\MSbar$ mass at a renormalization scale $\mu$ of order
$m_b$ still contains large corrections at low orders (see, Ref.~\cite{LSW,BBB}
and some examples in Sec.~V).  On the other hand, the perturbation series
written in terms of the $\MSbar$ mass at a low scale has a better behavior. 
But such a choice still does not remove the quark mass uncertainty in the decay
rates. 

The expansion of the $\Upsilon$ mass also becomes better behaved when written
in terms of the $\MSbar$ mass at a low scale (of order the Bohr radius,
$m\alpha_s$), rather than in terms of the pole mass.   This fact is  important
for extracting the short distance masses of the bottom and top quarks from the
$\Upsilon$ system and from the location of the peak of the $1S$ $t\bar t$
resonance in the total cross section for $t\bar t$ production in $e^+e^-$
annihilation close to threshold.  One might think that the numerical results
obtained in the upsilon expansion are equivalent to the result obtained by
extracting a short-distance $b$ quark mass from the $\Upsilon$ mass and then
using it to predict $B$ decay rates. Due to the fact that only the first few
terms in the perturbative expansions are known, this is not the case because in
this procedure the information on the correlations between the short-distance
mass, the strong coupling constant and the renormalization scale is lost. The
upsilon expansion, which completely eliminates the intermediate step of using
any quark mass takes full advantage of these correlations and leads to smaller
uncertainties.

The potential between two static quarks in a color singlet state,
$V(r)=-C_F\,g_s^2/r$, also has renormalon ambiguities.  
The Borel transform is given by~\cite{ugo}
\begin{equation}\label{borelpot}
\widetilde V(r,u) = - {4C_F e^{-Cu}\over\beta_0}\, \frac1r\, (\mu r)^{2u}\,
  {\Gamma(1/2+u)\,\Gamma(1/2-u)\over\Gamma(2u+1)} .
\end{equation}
The leading renormalon ambiguity of $V$ also comes from the $u=1/2$ pole,
\begin{equation}\label{delpot}
  \Delta V = {4C_F\over\beta_0}\, e^{-C/2} \lqcd \,.
\end{equation}
It follows from Eqs.~(\ref{borelmass}) and (\ref{borelpot}) that the order
$\lqcd$ ambiguity cancels in $2m_b+V$~\cite{andre,beneke}.

Equation~(\ref{borelpot}) also explains the mismatch between the power of
$\alpha_s$ and $\epsilon$ in Eq.~(\ref{schemups}), which may seem somewhat
unnatural at first sight.  The coefficient of $\alpha_s^{n+2}\beta_0^n$ in
Eq.~(\ref{schemups}) is given by the $\Upsilon$ matrix element of the
coefficient of $u^n$ in the expansion of $\widetilde V(u)$ in
Eq.~(\ref{borelpot}) about $u=0$, multiplied by $n!$. For large enough $n$, up
to corrections suppressed by at least one power of $n$, the terms in this
expansion of  $\widetilde V(r,u)$ are proportional to
\begin{equation}
{(2u)^n \over r}\,
\sum_{p=0}^n {\left[ \ln(\mu r) \right]^p \over p!} \simeq \mu\,(2u)^n.
\end{equation}
This explains why in the BLM approximation and in higher orders in 
$\epsilon$, the terms in Eq.~(\ref{schemups}) are generically of the 
form\footnote{
The actual coefficients of the logarithms in Eq.~(\ref{schemups}) 
differ from those displayed in Eq.~(\ref{genericexpansion}) because of
multiple insertions of the corrections to the pure Coulomb potential, 
$- C_F\alpha_s/r$.
The contributions arising from multiple insertions, however, only constitute
corrections suppressed by at least one power of $n$.   
} 
\begin{equation}
n!\,\alpha_s^{n+2} \beta_0^n
\left({\ell^n\over n!}+{\ell^{n-1}\over (n-1)!} +\ldots+1\right),
\label{genericexpansion}
\end{equation}
since the $\Upsilon$ matrix element of $\ln (\mu r)\sim \ell$.
For large $n$,
\begin{equation}
\left({\ell^n\over n!}+{\ell^{n-1}\over (n-1)!}
+\ldots+1 \right) \to \exp\left(\ell\right) ={ \mu\over m_b\alpha_s C_F},
\end{equation}
reducing the difference between the powers of $\alpha_s$ and
$\beta_0$ to one, thus eliminating the mismatch in the large order behaviors of
Eqs.~(\ref{schemrate}) and (\ref{schemups}).  This has to happen since
$m_\Upsilon$ is a physical quantity, so the renormalon ambiguities must cancel
in Eq.~(\ref{schemups}) between $2m_b$ and the potential plus kinetic
energies~\cite{andre,beneke}.

If NRQCD exists as a well defined low energy effective field theory, then the
$\lqcd^3 r^2$ renormalon in the potential corresponding to the $u=3/2$ pole in
Eq.~(\ref{borelpot}) must cancel against the leading renormalon in the kinetic
energy contribution to the $\Upsilon$ mass.  While this has not been shown
explicitly, it is quite conceivable that this happens.
Studying higher order renormalons is less interesting as there are
certainly nonperturbative contributions of order $\lqcd^4/(m\alpha_s)^3$ to
$m_\Upsilon$ (e.g., from the vacuum expectation value of
$G_{\mu\nu}^2$~\cite{Gmn2,Leutwyler1,Voloshin1}), and so  higher order
renormalon effects are of the same order as other nonperturbative effects which
are known to exist.

\subsection{Infrared sensitivity at low orders in $\epsilon$}

One can study the infrared sensitivity of Feynman diagrams by introducing a
fictitious infrared cutoff $\lambda$ of order $\lqcd$. The infrared sensitive
terms are nonanalytic in $\lambda^2$, such as $(\lambda^2)^{n/2}$ or
$\lambda^{2n}\ln\lambda^2$. These terms arise from the low-momentum part of
Feynman diagrams. This is similar to what occurs in chiral perturbation theory,
where the nonanalytic terms in the quark mass (or equivalently, the pion
mass-squared) come from the low-momentum behavior of Feynman diagrams. Diagrams
which are more infrared sensitive, i.e., have contributions $(\lambda^2)^{n/2}$
or $\lambda^{2n}\ln\lambda^2$ for small values of $n$, should have larger
nonperturbative contributions. In particular, linear infrared sensitivity,
i.e., terms of order $\lambda$, are a signal of $\lqcd$ effects, quadratic
sensitivity, i.e., terms of order $\lambda^2\ln \lambda^2$ are a signal of
$\lqcd^2$ effects, etc.

In the following we show that the linear infrared sensitivity present
separately in the expansion of the $B$ decay rate and in the expansion of the
$\Upsilon$ mass in terms of the $b$ quark mass cancel at least to order
$\epsilon^2$ when they are combined using the upsilon expansion.  (At order
$\epsilon$ this is equivalent to the cancellation of the leading renormalon
ambiguity in the BLM approximation, discussed previously.) The argument is
essentially a combination of those in recent papers by Beneke~\cite{beneke},
and by Sinkovics, Akhoury, and Zakharov~\cite{SAZ}. As far as the cancellation
of infrared sensitive contributions in the perturbative expansion of $B$ decay
is concerned, the upsilon expansion is therefore equivalent to a scheme where a
short-distance mass like the $\MSbar$ mass is used.

The proof proceeds in two steps. First one notes that the linear IR sensitivity
cancels in inclusive decay widths of a heavy quark when it is expressed in
terms of an infrared safe short distance mass (such as the $\MSbar$ mass). To
leading order in $\alpha_s$ this was shown in
Refs.~\cite{rencan1,rencan2,rencan3,ren1}. To order $\alpha_s^2$, where
non-Abelian contributions first occur, the cancellation of the linear IR
sensitivity was shown only recently in Ref.~\cite{SAZ}. This cancellation of
the linear IR sensitivity is expected to hold at higher orders in $\alpha_s$ as
well, but the demonstration of this appears highly non-trivial. The second step
is to show the cancellation of the linear IR sensitivity when the $\Upsilon$
mass is expressed in terms of an infrared safe short distance mass.  This was
done by Beneke~\cite{beneke}, and we will repeat some of his arguments below. 
The reason is to explain why it is imperative to assign
one less power of $\epsilon$ to each term in Eq.~(\ref{upsmass}) than the power
of $\alpha_s$ even at low orders in $\epsilon$.

\begin{figure}
\epsfxsize=10cm
\hfil\epsfbox{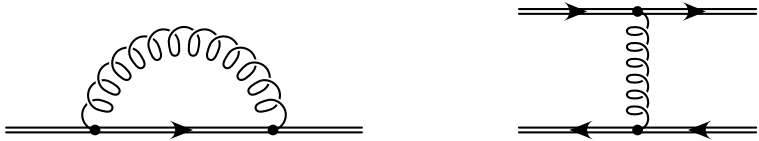}\hfil
\caption{One-loop correction to the heavy quark self energy, and the tree-level
contribution to the $Q\bar Q$ potential.
\label{fig:hqselfenergy}}
\end{figure}

Let us first consider the leading contribution to the heavy quark self energy,
shown in Fig.~\ref{fig:hqselfenergy}.\footnote{The argument in this and the
next paragraph are directly from Beneke~\cite{beneke}.}
Since we are only interested in the infrared behavior, we may calculate the
self energy diagram in terms of the bare mass, $m_0$, which differs from the
$\MSbar$ mass only by ultraviolet subtractions.  The self energy is given by
\begin{equation}
-i \Sigma = - C_F\, g_s^2 \int {{\rm d}^4q\over(2\pi)^4}\,
  {\gamma_\mu(\psla+\qsla+m_0)\gamma^\mu \over [(p+q)^2-m_0^2+i\varepsilon]
  (q^2+i\varepsilon)} \,.
\end{equation}
To obtain the leading infrared contribution to the mass shift coming from the
region of small loop momentum, we can set $p=m_0 v$, where $v$ is the 
four-velocity of the heavy quark ($v^2=1$).  This gives
\begin{equation}
\delta m = -iC_F\, g_s^2 \int {{\rm d}^4q\over(2\pi)^4}\,
  {1 \over (v\cdot q+i\varepsilon) (q^2+i\varepsilon)} \,.
\end{equation}
The integration over $q^0$ is simplest by closing the contour in the upper half
plane, yielding
\begin{equation}\label{dmassIR}
\delta m = \frac12 C_F\, g_s^2 \int {{\rm d}^3 {\mathbf q}\over(2\pi)^3}\,
  {1\over {\mathbf q}^{2} } \,.
\end{equation}
Of course, this formula is only valid in the infrared.  Restricting the range
of integration to $|{\mathbf q}|<\lambda$, where $\lambda$ is of order $\lqcd$,
it gives $\delta m\sim\alpha_s\lambda$. 

Let us next consider the heavy quark potential.  It was shown in
Ref.~\cite{beneke} that the potential in momentum space is free of linear IR
sensitivity.  The linear IR sensitivity of the contribution of the potential to
$m_\Upsilon$ arises in a momentum space calculation from taking its matrix
element.  This occurs because the momentum space wave function $\sim 1 /(1 +
a^2q^2)^2$ does not suppress the low momentum contribution of $V(q)$.  In
coordinate space, the wave function $\sim \exp(-r/a)$ is exponentially small at
long distances, and the IR sensitivity of the contribution of the potential to
$m_\Upsilon$ arises from Fourier transforming $V(q)$ into coordinate space to
obtain $V(r)$.  Therefore, the linear infrared sensitivity of the potential
energy can be isolated by restricting the range of integration in the Fourier
transform~\cite{beneke}
\begin{equation}\label{dpot}
\delta V = \int_{|{\mathbf q}|<\lambda} {{\rm d}^3 {\mathbf q}\over(2\pi)^3}\,
  e^{i{\mathbf q\cdot r}}\, V({\mathbf q}) \,.
\end{equation}
Choosing $\lambda$ of order $\lqcd$, much smaller than $1/a$, the $e^{i{\mathbf
q \cdot r}}$ factor can be replaced by 1, and the IR sensitive term is a
constant  times $\alpha_s \lambda$.\footnote{ The assumption $\lqcd \ll 1/a$ is
essential for the argument to hold. For the $\Upsilon(1S)$ ground state this
assumption is valid. }   Since this contribution to the potential is
independent of $r$, its $\Upsilon$ matrix element is simply itself,
\begin{equation}
\Bigl\langle \Upsilon \Bigr|\, \alpha_s \lambda \, \Bigl| \Upsilon
\Bigr\rangle \sim  \alpha_s\lambda
\,.
\label{alphaslambda}
\end{equation}
{}From Eq.\ (\ref{alphaslambda}) it is evident that the infrared sensitive
contribution in the $\Upsilon$ matrix element of the heavy quark potential is
of one lower order in $\alpha_s$ than the $\Upsilon$ matrix element of the
heavy quark potential itself in the limit $\lambda\to 0$,
\begin{equation}
\Bigl\langle \Upsilon \Bigr| \,\frac{\alpha_s}{r}  \,\Bigl| \Upsilon
\Bigr\rangle \sim  m_b\,\alpha_s^2
\,.
\end{equation}
This mismatch arises because the $\Upsilon$ matrix element of $1/r$  is of order
the inverse Bohr radius $1/a \sim m_b\alpha_s$.  Consequently, it is the infrared
sensitive contribution in the order $\alpha_s^2$ term in the perturbative
series for $m_\Upsilon$ in Eq.\ (\ref{upsmass}) which cancels the infrared
sensitive contribution of the order $\alpha_s$ term in the perturbative series
for inclusive $B$ decay rates in  Eq.~(\ref{schemrate}) (or equivalently 
for the $b$
quark pole mass). It was shown in Ref.~\cite{beneke} that this cancellation of
the linear IR sensitivity occurs similarly to all orders in
$\alpha_s$.\footnote{For Beneke's all order argument
to hold one has to assume that the heavy quark potential does not contain any
infrared diverences (or equivalently infrared sensitivity to scales below the
inverse Bohr radius) to all orders in $\alpha_s$. It was argued in 
Ref.~\cite{ADM} that this assumption may not hold beyond order $\alpha_s^3$.
To order $\alpha_s^3$, i.e. to order $\epsilon^2$, 
this assumption is proven by explicit 
calculations~\cite{Fischler1,Billoire1,Peter1}. }

Thus it is shown that using the upsilon expansion for $B$ decay rates 
eliminates all linear infrared sensitivity at least to order $\epsilon^2$. 
This is the order to which most $B$ decay rates have been computed, and the
order to which we will present results in Secs.~IV and V.  While we cannot
prove that this cancellation is present at all orders in perturbation theory,
this is likely in view of the fact that it holds both at large orders in the
bubble summation approximation and in the first two orders including
non-Abelian contributions. Consequently, the upsilon expansion is insensitive
to any order $\lqcd$ ambiguity in the $b$ quark mass which is present in
perturbation theory.

\subsection{Nonperturbative corrections to the $\Upsilon$ mass}

An important theoretical uncertainty in applying the  upsilon expansion is the
size of nonperturbative corrections to the $\Upsilon$ mass in
Eq.~(\ref{upsmass}).  The dynamics of the $\Upsilon$ system can be described
using NRQCD~\cite{NRQCD}.  The leading nonperturbative corrections to
$m_\Upsilon$ arise from matrix elements in the $\Upsilon$ of $H_{\rm light}$,
the Hamiltonian of the light degrees of freedom. In $B$ mesons, the leading
nonperturbative correction to the $B$ meson mass is due to the matrix element
of $H_{\rm light}$, which is the $\bar\Lambda$ parameter of order $\lqcd$. The
$\lqcd$ dependence is different for the $\Upsilon$. $H_{\rm light}$ is the
integral of a local Hamiltonian density,
\begin{equation}
H_{\rm light}=\int\! {\rm d}^3x\, {\mathcal H}_{\rm light}(x).
\end{equation}
The radius of the $\Upsilon$ is $a\sim1/(m_b \alpha_s)$, so the matrix element
of $H_{\rm light}$ is of order $a^3\lqcd^4$, by dimensional analysis. (Note
that the matrix element of  $H_{\rm light}$ is of order $\lqcd^4$, not
$m_b^4$.  Terms that grow with $m_b$ can be treated using NRQCD perturbation
theory.) Using $1/a\sim 1\,$GeV, and $\lqcd \sim 350\,$MeV, of order a
constituent quark mass, gives a nonperturbative correction of $15\,$MeV. Using
instead $\lqcd\sim 500\,$MeV gives a correction of $60\,$MeV. We will use
$100\,$MeV as a conservative estimate of the nonperturbative contribution to
$m_\Upsilon$.

For large enough quark mass, the dominant nonperturbative corrections to the
quarkonia masses are believed to come from the gluon condensate, $\langle
\alpha_s G_{\mu\nu}^2 \rangle$. Its contribution to the mass of the $n^{\rm
th}$ $S$-wave state as calculated in Refs.~\cite{Leutwyler1,Voloshin1} is
\begin{equation}\label{nonpert}
\delta m^{(n)} = m_b\,
  {n^6 e_n \pi \langle \alpha_s G_{\mu\nu}^2 \rangle \over
  (m_b C_F \widetilde\alpha_s)^4 } \,,
\end{equation}
where $\alpha_s$ in the denominator should be evaluated at the scale of the
order of the Bohr radius of the $n^{\rm th}$ state, $\widetilde\alpha_s \sim
\alpha_s(a^{-1}/n^2)$, and $e_1=1.41$, $e_2=1.59$, etc., are numerical
constants.  Values of the gluon condensate in the literature vary in the range
$\langle \alpha_s G_{\mu\nu}^2 \rangle \sim (0.05\pm0.03)\,{\rm GeV}^4$, and so
the uncertainties in Eq.~(\ref{nonpert}) are sizable. In Ref.~\cite{PY},
Eq.~(\ref{nonpert}) is estimated to give a 60~MeV contribution to the
$\Upsilon(1S)$ mass. Equation~(\ref{nonpert}) seems to be enhanced by
$1/\alpha_s$ compared to our naive estimate of $a^3 \lqcd^4$ above. This is due
to the fact that $\langle \alpha_s G_{\mu\nu}^2  \rangle$ and not $\langle
G_{\mu\nu}^2 \rangle$ is renormalization group  invariant. In our estimate the
operator $\mathcal{H}_{\rm light}$ is renormalized at the scale $\mu=a^{-1}$.
The renormalization group scaling from $\mu=a^{-1}$ to a low scale $\mu \sim
1$~GeV gives an additional factor $\alpha_s(1 \hbox{GeV})/\alpha_s(a^{-1})$,
which is parameterically of order $1/\alpha_s(a^{-1})$. Since $a^{-1}\sim
1$~GeV, we have not included this factor in our estimate.

Our equations depend on the $\Upsilon$ mass, and make no reference to the wave
function of the $\Upsilon$.  In fact, there are much larger nonperturbative
contributions to the $\Upsilon(1S)$ wave function, which nevertheless, do not
affect the energy of the state~\cite{Voloshin1}. The reason is that the energy
can be obtained in terms of the matrix element of the NRQCD effective
Hamiltonian, whereas the wave function has no simple operator expansion. The
situation is similar to other well-studied processes, such as the cross-section
for $e^+e^-\to \hbox{hadrons}$, where the total cross-section has an OPE
description and much smaller nonperturbative corrections than other non-OPE
observables such as event shapes~\cite{event}.

{}From the results presented in Ref.~\cite{prl}, one can see three
indications that the nonperturbative
corrections to Eq.~(\ref{upsmass}) may be smaller than typical expectations based on
Eq.~(\ref{nonpert}).  First, the sign of $\langle \alpha_s G_{\mu\nu}^2
\rangle$ is positive, and a large value for it increases $|V_{cb}|$ as
extracted from the inclusive semileptonic $B$ width.  Thus, for a very large
nonperturbative correction to $m_\Upsilon$ (say, above $500\,$MeV or so), the
consistency of the inclusive and exclusive $|V_{cb}|$ determinations is
significantly worse.  Second, in Ref.~\cite{prl} we determined $|V_{cb}|$ using
both the mass of the $\Upsilon(1S)$ and $\Upsilon(2S)$.  Eq.~(\ref{nonpert})
shows that nonperturbative corrections to the $n^{\rm th}$  $S$-wave state
scale as $n^6$.  If we assume that Eq.\ (\ref{nonpert}) holds at least
approximately also for the size of nonperturbative effects for higher radial
excitations of the $\Upsilon$, the nonperturbative contributions  would be  64
times larger for the $2S$ than for the $1S$.  The fact that the resulting
values of $|V_{cb}|$ coincide at the 5\% level at order $\epsilon^2$~\cite{prl}
indicates that there cannot be a large nonperturbative correction to the mass
of the $\Upsilon(1S)$. Third, the ``most nonperturbative'' that an $\Upsilon$
can be is for it to look like a $B-\bar B$ bound state, which occurs just at
threshold for decay into two mesons. The nonperturbative contribution to the
$B+\bar B$ mass is $2\bar\Lambda$, and is of order $1$~GeV. Assuming  again
that this is  approximately  the nonperturbative contribution for $n=3$, one
finds that the $\Upsilon(1S)$ state has a nonperturbative contribution of order
$1\ \hbox{GeV}/3^6\sim 1.4$~MeV, an absurdly small value.

Finally, note that one cannot separate perturbative and nonperturbative
contributions to a physical quantity, since the perturbation series is a
divergent series.\footnote{This separation is not possible even if an infrared
cutoff is introduced, since the perturbation series is still divergent. 
The $n!$ growth in the coefficients due to infared renormalons is
removed, but there are other sources of factorial growth in the coefficients,
such as the number of diagrams at $n^{\rm th}$ order~\cite{factgrowth}.} One can treat the
perturbation series as an asymptotic series, and sum the terms until they no
longer decrease in magnitude. The resulting difference from the complete answer
(which is not known) can be called the nonperturbative contribution. If the
perturbation series is poorly behaved, so that one can keep only a few terms in
the asymptotic expansion, then one would expect that the error in the
asymptotic expansion is large---i.e., that nonperturbative effects are also
large. To make this more precise, consider the $e^+e^-$ total cross-section
into hadrons at $s=Q^2$, in the limit of massless quarks. By dimensional
analysis, the cross-section must have the form
\begin{equation}
Q^2\, \sigma(s=Q^2) = f\left(\alpha_s(Q)\right),
\end{equation}
The dimensionless function $f(\alpha_s(Q))$ includes both perturbative and
non-perturbative effects, and has an asymptotic series expansion in
$\alpha_s$. For an asymptotic expansion,
the best approximation to $f$ is given by summing the series till the terms
start to increase. The error at the minimum term, which is typically of the
form
\begin{equation}
e^{-c/\alpha_s(Q)} \sim \left({\lqcd \over Q}\right)^{c \beta_0/4\pi}
\end{equation}
can be considered the nonperturbative contribution to $f$.

The terms in the $\alpha_s$ expansion for the $\Upsilon$ mass  in Eq.\
(\ref{upsmass}) start increasing already at order $\epsilon^2$. Using the order
$\epsilon$ term as an estimate of the error in the asymptotic series gives
$100$~MeV, which can be taken as an estimate for the size of nonperturbative
contributions. When one combines $m_\Upsilon$ with $B$ decay rates, the
perturbation series is much better behaved, and one expects a much smaller
nonperturbative contribution to $B$ decay rates expressed in terms of
$m_\Upsilon$. In other words, the  dominant part of the  large nonperturbative
contributions to $m_\Upsilon$ are closely related to the $\lqcd$ uncertainty in
the pole mass. What is relevant for our analysis is the nonperturbative
contributions to $m_\Upsilon$ after these pole mass effects have been removed.
The residual contributions are expected to be much smaller, of order
$\sim\lqcd^4/(m_b \alpha_s)^3$, and formally of the same order as $1/m_b^3$
corrections which have been neglected in the inclusive decay rates. It is these
residual nonperturbative effects for which we have used the estimate of
100~MeV.

To summarize, we think that the size of nonperturbative corrections to
$m_\Upsilon$ relevant for our analysis are hardly known at the order of
magnitude level (the conventional estimates yield hundreds of MeV, but our
arguments above indicate that they could equally be tens of MeV only). The best
way of determining them is  directly from the experimental data. We would like
to find  $B$ decay observables sensitive to $m_\Upsilon$ for which
parameter-free predictions can be made.   The best candidates are shape
variables in inclusive decays or form factor ratios in exclusive decays. The
reason is that observables like total inclusive rates are only moderately
sensitive to $m_\Upsilon$, and a possible discrepancy between experiment and
theory could always be interpreted as a different value for a CKM angle, as
opposed to constraining nonperturbative corrections to $m_\Upsilon$.  The most
promising observables are the ones which are the most sensitive to
$\bar\Lambda$: the mean photon energy in $B\to X_s\gamma$, and the form factor
ratio $R_1(w) = h_V(w) / h_{A_1}(w)$ in exclusive $B\to D^* e\bar\nu$ decay.
These are discussed in Secs.~V.B and IV, respectively.

\section{Exclusive Decays}

Exclusive semileptonic $B\to D^* e\bar\nu$ decays depend on four form factors,
\begin{eqnarray}\label{formf}
{\langle D^*(v',\epsilon)|\, V^\mu\, |B(v)\rangle \over \sqrt{m_{D^*}\,m_B}}
  &=& i\, h_V\, \varepsilon^{\mu\alpha\beta\gamma} 
  \epsilon^*_\alpha v'_\beta v_\gamma \,, \nonumber\\*
{\langle D^*(v',\epsilon)|\, A^\mu\, |B(v)\rangle \over \sqrt{m_{D^*}\,m_B}}
  &=& h_{A_1} (w+1)\, \epsilon^{*\mu} 
  - (h_{A_2} v^\mu + h_{A_3} v'^\mu)\, (\epsilon^*\cdot v) \,,
\end{eqnarray}
where $w=v \cdot v^\prime$.
In the infinite mass limit $h_{A_2}=0$ and the other three are equal to the
Isgur-Wise function~\cite{HQS}. One linear combination of the form factors is
not measurable when the lepton masses are neglected.  It is
conventional~\cite{physrep} to define two ratios of the three measurable form
factors
\begin{equation}\label{R12def}
R_1(w) = {h_V(w) \over h_{A_1}(w)}\,, \qquad
R_2(w) = {h_{A_3}(w) + (m_{D^*}/m_B) h_{A_2}(w) \over h_{A_1}(w)}\,.
\end{equation}
Here we shall concentrate on the prediction for $R_1(1)$ at order 
$(\epsilon^2)_{\rm BLM}$ in the upsilon expansion.  Using the notation and
results of Ref.~\cite{physrep}, the theoretical prediction for $R_1$ is
\begin{equation}\label{R1exp}
R_1(1) = {C_1\over C_1^5} + {\bar\Lambda\over 2m_c} 
  + {\bar\Lambda\over 2m_b}\, [1-2\eta(1)] + \ldots \,, \qquad
\end{equation}
where the elipses denote the known order $\alpha_s/m_{c,b}$ and unknown order
$1/m_{c,b}^2$ corrections.  If the QCD sum rule prediction
$\eta(1)=0.6\pm0.2$~\cite{eta} is approximately correct, then the
$\bar\Lambda/m_b$ correction is much smaller than the $\bar\Lambda/m_c$ term. 
One can avoid relying on a theoretical estimate for $\eta(1)$,
since the deviation of $R_2$ from unity
measures $\eta(w)$ directly if the $\chi_2$ subleading Isgur-Wise function is
neglected.  Since $\chi_2$ parameterizes $1/m_{c,b}$ corrections due to
insertion of the chromomagnetic operator in the Lagrangian, it is expected to
be small and a QCD sum rule calculation supports this~\cite{chi2}.  Neglecting
$\chi_2$ seems to be less model dependent than relying on the previously
mentioned result for $\eta(1)$. The relation between $R_2(1)$ and $\eta(1)$ 
when $\chi_2$ is neglected is
\begin{equation}
\eta(1) = [1-R_2(1)]\,{2m_c \over \bar\Lambda (1+3z)},
\end{equation}
where $z=m_c/m_b$.

The perturbative correction $C_1^5 = \eta_A$ is known to order $\alpha_s^2
\beta_0$~\cite{etaBLM},
\begin{equation}
C_1^5 = \eta_A = 1 + {\bar\alpha_s \over 4\pi}\, C_F [\phi(z)-2] +
  \bigg( {\bar\alpha_s \over 4\pi} \bigg)^2 \beta_0 C_F 
  \bigg[ \frac56\, \phi(z) - \frac73 \bigg] \,,
\end{equation}
where $\bar\alpha_s = \alpha_s(\sqrt{m_c m_b}) \simeq 0.27$ and
\begin{equation}
\phi(z) = -3 \, {1+z \over 1-z} \ln(z) -6.
\end{equation}
{}From results given
in Ref.~\cite{rencan3}, we can obtain the coefficient $C_1$ to
order $\alpha_s^2\beta_0$
\begin{equation}
C_1 = 1 + {\bar\alpha_s \over 4\pi}\, C_F [\phi(z)+2] +
  \bigg( {\bar\alpha_s \over 4\pi} \bigg)^2 \beta_0 C_F 
  \bigg[ \frac{13}6\, \phi(z) + \frac{25}3 \bigg] \,.
\end{equation}
To combine these results with Eq.~(\ref{upsmass}), it seems simplest to 
reexpress $\bar\alpha_s$ in terms of $\alpha_s(m_b)$.  This yields
\begin{equation}\label{Cratio}
{C_1 \over C_1^5} = 1 + {\alpha_s \over \pi}\, C_F +
  \bigg( {\alpha_s \over 4\pi} \bigg)^2 \beta_0 C_F\, 8
  \bigg( \frac13 - {\ln z \over 1-z} \bigg) \,.
\end{equation}
Thus we obtain at order $(\epsilon^2)_{\rm BLM}$ in the upsilon expansion
\begin{equation}\label{R1pred}
R_1(1) = 1.21 + 0.07\epsilon + 0.01_{\rm BLM}\epsilon^2
  + 0.01\lambda_1/{\rm GeV}^2 + [0.31R_2(1)-0.25] \,.
\end{equation}
The quantity $\lambda_1$ appears in this result because we used
Eq.~(\ref{mbmc}) to eliminate the dependence on the charm quark mass. The last
term in square brackets is the $\bar\Lambda/m_b$ contribution
in Eq.~(\ref{R1exp}), and we have assumed that it is small, so that terms of
order $\epsilon$ or $\lambda_1$ from this contribution have been omitted.
A $100\,$MeV nonperturbative contribution to
$m_\Upsilon$ changes the prediction for $R_1$ by $\pm 0.021$, which is about
twice the order $(\epsilon^2)_{\rm BLM}$ correction.  Probably the largest
uncertainty is due to the neglected $1/m_{c,b}^2$ corrections~\cite{FaNe},
which need to be better understood before Eq.~(\ref{R1pred}) may be trusted at
the percent level.

One can compare Eq.~(\ref{R1pred}) with the conventional treatment. The size of
the order $\alpha_s^2\beta_0$ terms is usually quoted in terms of the BLM
scale. The BLM scale of $C_1^5 = \eta_A$ is $0.51\sqrt{m_c m_b}$ and that of
$C_1+C_2+C_3 = \eta_V$ is $0.92\sqrt{m_c m_b}$~\cite{etaBLM}.  Surprisingly,
the BLM scale of $C_1$ is much smaller, $0.16\sqrt{m_c m_b}$, and that of
$C_1/C_1^5$ in Eq.~(\ref{Cratio}) is only slightly larger, $0.13m_b$. These low
BLM scales show that the conventional perturbation series for $R_1$ is poorly
behaved. The perturbation series in Eq.~(\ref{R1pred}) cannot be converted into
a BLM scale since it mixes different orders in $\alpha_s$. However, to compare
with conventional results one can define an effective BLM scale by using the
ratio of the $(\epsilon^2)_{\rm BLM}$ and $\epsilon$ terms, just as the
conventional treatment used the ratio of the $(\alpha_s^2)_{\rm BLM}$ and
$\alpha_s$ terms. The effective BLM scale of Eq.~(\ref{R1pred}) is $0.89 m_b$,
which is large and allows for a reliable expansion.

The present experimental results are~\cite{CLEO}
\begin{equation}\label{R12data}
R_1 = 1.18 \pm 0.30 \pm 0.12 \,, \qquad
R_2 = 0.71 \pm 0.22 \pm 0.07 \,.
\end{equation}
This is consistent with Eq.~(\ref{R1pred}), although the errors are sizable. 
If $R_{1,2}(1)$ will be measured in the future with one (or at most a few)
percent errors, then it may be possible to constrain nonperturbative
contributions to $m_\Upsilon$ at the $100\,$MeV level, which would be very
interesting.  

To summarize, Eq.~(\ref{R1pred}) predicts the form factor ratio $R_1$
independent of $\bar\Lambda$ and the quark masses.  If this is tested and works
at the few percent level, then we can make better and still model independent
predictions for $B$ decay form factors in general, including the ones to the
orbitally excited mesons $D_1$ and $D_2^*$~\cite{llsw}.  Even if
nonperturbative contributions to $m_\Upsilon$ cannot be constrained too well,
our results show that expressing $B$ decay rates in terms of $m_\Upsilon$ is  a
useful and theoretically consistent parameterization.

\section{Inclusive $B$ decay rates}

In this section, we apply the upsilon expansion to inclusive $B$ decays. The
$b$ quark decay rate has long been known for arbitrary masses of the four
fermions participating in the weak decay, both at tree level~\cite{CPT} and at
order $\alpha_s$~\cite{HoPa}.  The order $\lqcd^2/m_b^2$ nonperturbative
corrections will be taken into account for all decay modes, however, $\alpha_s$
corrections to these will be neglected. The numerical results below correspond to the choice $\mu=m_b$
and $\alpha_s(m_b)=0.22$, and $\mu$-dependence will be discussed as well. 

Nonleptonic decays are more complicated than semileptonic decays. It is less
clear whether local duality holds in nonleptonic decays, since there is no
smearing variable~\cite{PQW,LM}. There are also corrections not present in 
semileptonic decay
because the four-quark operators run in the effective theory below the weak
scale. While the complete series of leading~\cite{Alt} and
subleading~\cite{Ball} logarithms have been summed, we cannot consistently use
these results, since that would involve summing series of the form $[\alpha_s
\ln(m_W/m_b)]^n$ and $\alpha_s [\alpha_s\ln(m_W/m_b)]^n$ (where $\ln(m_W/m_b)
\sim 2.8$) without summing $\alpha_s^{n+1} \beta_0^n$ (where $\beta_0 \sim
9$).  Instead, we follow an approach somewhat similar to Ref.~\cite{LLSS} based
on the observation that the largest corrections arise at low orders in
$\alpha_s$.  This is quite reasonable, as the order $[\alpha_s \ln(m_W/m_b)]^2$
term dominates the sum of leading logs (there is no order $\alpha_s
\ln(m_W/m_b)$ term due to color conservation), and the order $\alpha_s$ term
dominates the sum of next-to-leading logs.  For $b\to c$ nonleptonic decays we
include the order $\alpha_s$ correction, and at order $\alpha_s^2$ the terms
containing $[\ln(m_W/m_b)]^2$, $\ln(m_W/m_b)$, and $\beta_0$.\footnote{It was
argued that other order $\alpha_s^2$ terms might also be unexpectedly large due
to Coulomb enhancement~\cite{MBV}.} For $b\to u$ nonleptonic decays we include
the order $\alpha_s$ and order $[\alpha_s \ln(m_W/m_b)]^2$ corrections only.  

\subsection{Semileptonic $B$ decays}

Implications of our method for semileptonic $B$ decays were studied in
Ref.~\cite{prl}, and here we repeat the results briefly for completeness. 
Estimates of the small non-BLM contributions to the decay rates at order
$\alpha_s^2$, which were taken into account in Ref.~\cite{prl}, will be
neglected here.  For massless leptons, the $\alpha_s$ correction to free quark
decay is known analytically~\cite{Yossi}, and the order $\alpha_s^2 \beta_0$
correction has been determined in Ref.~\cite{LSW}.

\subsubsection{$B\to X_u\,e\,\bar\nu$}

The $B\to X_u e\bar\nu$ decay rate at order $(\epsilon^2)_{\rm BLM}$ in the
upsilon expansion is~\cite{prl}
\begin{eqnarray}\label{buups}
\Gamma(B\to X_u e\bar\nu) = {G_F^2 |V_{ub}|^2\over 192\pi^3}
  \bigg({m_\Upsilon\over2}\bigg)^5\, \bigg[ 1 &-& 0.115\epsilon 
  - 0.035_{\rm BLM} \epsilon^2 \nonumber\\*
&-& {9\lambda_2 - \lambda_1 \over 2(m_\Upsilon/2)^2} + \ldots \bigg] ,
\end{eqnarray}
The perturbation series, $1 - 0.115\epsilon - 0.035_{\rm BLM}\epsilon^2$, is
far better behaved than the series in terms of the pole mass, $1 - 0.17\epsilon
- 0.13_{\rm BLM}\epsilon^2$, or the series expressed in terms of the $\MSbar$
mass at $\mu=m_b$, $1+0.30\epsilon+0.19_{\rm BLM}\epsilon^2$.  The uncertainty in the $B$
decay rate using Eq.~(\ref{buups}) is much smaller than that in
Eq.~(\ref{schemrate}), both because the perturbation series is better behaved,
and because the $\Upsilon$ mass is better known (and better defined) than the
$b$ quark mass.

Eq.~(\ref{buups}) yields a relation between $|V_{ub}|$ and the total
semileptonic $B\to X_u e\bar\nu$ decay rate with very small
uncertainty,\footnote{In Eqs.~(\ref{Vub}) and (\ref{Vcb}), we have included the
non-BLM terms of order $\epsilon^2$ as well (see Ref.~\cite{prl}).}
\begin{eqnarray}\label{Vub}
|V_{ub}| &=& (3.06 \pm 0.08 \pm 0.08) \times 10^{-3} \nonumber\\
&& \times \left( {{\cal B}(B\to X_u e\bar\nu)\over 0.001}
  {1.6\,{\rm ps}\over\tau_B} \right)^{1/2} ,
\end{eqnarray}
where we have used $\lambda_2 = 0.12\,{\rm GeV}^2$ and $\lambda_1 =
(-0.25\pm0.25)\,{\rm GeV}^2$. The first error is obtained by assigning an
uncertainty in Eq.~(\ref{buups}) equal to the value of the $\epsilon^2$ term
and the second is from assuming a $100\,$MeV uncertainty in
Eq.~(\ref{upsmass}).  The scale dependence of $|V_{ub}|$ due to varying $\mu$
in the range $m_b/2< \mu <2m_b$ is less than 1\%.  The uncertainty in
$\lambda_1$ makes a negligible contribution to the total error.  It is unlikely
that ${\cal B}(B\to X_u e\bar\nu)$ will be measured without significant
experimental cuts, for example, on the hadronic invariant mass~\cite{FLW}.  Our
method should reduce the uncertainties in such analyses as well.

\subsubsection{$B\to X_c\,e\,\bar\nu$}

The $B\to X_c e\bar\nu$ decay rate at order $(\epsilon^2)_{\rm BLM}$
is~\cite{prl}
\begin{eqnarray}\label{bcups}
\Gamma(B\to X_c e\bar\nu) = {G_F^2 |V_{cb}|^2\over 192\pi^3}
  \bigg({m_\Upsilon\over2}\bigg)^5\, 0.533 \Big[ 1 &-& 
  0.096\epsilon - 0.029_{\rm BLM}\epsilon^2 \nonumber\\
&-& (0.28\lambda_2 + 0.12\lambda_1)/{\rm GeV}^2 \Big] \,, 
\end{eqnarray}
For comparison, the perturbation series in this relation when written in terms 
of the pole mass is $1 - 0.12\epsilon - 0.07_{\rm BLM}\epsilon^2$. 
Equation~(\ref{bcups}) implies
\begin{eqnarray}\label{Vcb}
|V_{cb}| &=& (41.6 \pm 0.8 \pm 0.7 \pm 0.5) \times 10^{-3} \nonumber\\
&& \times \eta_{\rm QED} \left( {{\cal B}(B\to X_c e\bar\nu)\over0.105}\,
  {1.6\,{\rm ps}\over\tau_B}\right)^{1/2} ,
\end{eqnarray}
where $\eta_{\rm QED}\sim1.007$ is the electromagnetic radiative correction.
The uncertainties come from assuming an error in Eq.~(\ref{bcups}) equal to the
$\epsilon^2$ term, the $0.25\,{\rm GeV}^2$ error in $\lambda_1$, and a
$100\,$MeV error in Eq.~(\ref{upsmass}), respectively.  The second uncertainty
is reduced to $\pm0.3$ by extracting $\lambda_1$ from the electron spectrum in
$B\to X_c e\bar\nu$; see Eq.~(\ref{lambda1}).  The agreement of $|V_{cb}|$ with
other determinations (such as exclusive decays) is a check that nonperturbative
corrections to Eq.~(\ref{upsmass}) are indeed small.  

In Ref.~\cite{gremmetal} $\bar\Lambda$ and $\lambda_1$ were extracted from the
lepton spectrum in $B\to X_c e\bar\nu$ decay.  With our approach, there is no
dependence on $\bar\Lambda$, so we can determine $\lambda_1$ directly with
small uncertainty.  Considering the observable $R_1 = \int_{1.5{\rm GeV}} E_e
({\rm d}\Gamma / {\rm d}E_e) {\rm d}E_e \big/ \int_{1.5{\rm GeV}} ({\rm
d}\Gamma / {\rm d}E_e) {\rm d}E_e$, a fit to the same data yields
\begin{equation}\label{lambda1}
  \lambda_1 = (-0.27 \pm 0.10 \pm 0.04) \,{\rm GeV}^2 .
\end{equation}
The central value includes corrections of order $\alpha_s^2
\beta_0$~\cite{gremmetal2}.  The first error is dominated by $1/m_b^3$
corrections~\cite{gremmetal3}.  We varied the dimension-six matrix elements
between $\pm(0.5\,{\rm GeV})^3$, and combined their coefficients in quadrature
in the error estimate.  The second error is from assuming a $100\,$MeV
uncertainty in Eq.~(\ref{upsmass}).  The central value of $\lambda_1$ at tree
level or at order $\alpha_s$ is within $0.03\,{\rm GeV}^2$ of the one in
Eq.~(\ref{lambda1}).  This value of $\lambda_1$ can be used consistently in
theoretical expressions up to order $\alpha_s^2 \beta_0$~\cite{invisible}.

\subsubsection{$B\to X_c\,\tau\,\bar\nu$, $B\to X_u\,\tau\,\bar\nu$ and
$R_\tau$}

The order $\lqcd^2/m_b^2$ corrections are taken from~\cite{FLNN,tau}, and the
order $\alpha_s^2 \beta_0$ corrections from~\cite{LLSS}.  The $B\to
X_c\tau\bar\nu$ decay rate at order $(\epsilon^2)_{\rm BLM}$ is
\begin{eqnarray}\label{bctau}
\Gamma(B\to X_c\tau\bar\nu) = {G_F^2 |V_{cb}|^2\over 192\pi^3}
  \bigg({m_\Upsilon\over2}\bigg)^5\, 0.114 \Big[ 1 &-& 
  0.070\epsilon - 0.016_{\rm BLM}\epsilon^2 \nonumber\\
&-& (0.59\lambda_2 + 0.25\lambda_1)/{\rm GeV}^2 \Big] \,, 
\end{eqnarray}
For comparison, the perturbation series in this relation when written in terms 
of the pole mass is $1 - 0.097\epsilon - 0.064_{\rm BLM}\epsilon^2$. 

For the $B\to X_u\tau\bar\nu$ decay rate we find
\begin{equation}\label{butau}
\Gamma(B\to X_u \tau\bar\nu) = {G_F^2 |V_{ub}|^2\over 192\pi^3}
  \bigg({m_\Upsilon\over2}\bigg)^5\, 0.361 \Big[ 1 -
  0.08\epsilon - (0.34\lambda_2 - 0.02\lambda_1)/{\rm GeV}^2 \Big] \,, 
\end{equation}
For comparison, the order $\epsilon$ correction written in terms of the pole
mass is $1 - 0.16\epsilon$.

The ratio $R_\tau = {\cal B}(B\to X_c\tau\bar\nu) / {\cal B}(B\to X_c
e\bar\nu)$ is interesting because it directly probes lepton mass effects, and
it can be predicted precisely in the standard model.  The largest uncertainty
in the prediction is $\lambda_1$, and varying it in the range given in
Eq.~(\ref{lambda1}) yields $0.220 < R_\tau < 0.227$.\footnote{ For this
calculation we have set ${\cal B}(B\to X_c e\bar\nu)=10.5\%$, and neglected its
uncertainty.} Thus, in the standard model, measurements of $R_\tau$ can be used
to constrain $\lambda_1$~\cite{FLNN,YZ,FLS}.  The most recent ALEPH data,
${\cal B}(B\to X_c\tau\bar\nu) = (2.72\pm0.20\pm0.27)\%$~\cite{Alephtau},
implies at the $1\sigma$ level $\lambda_1 < -0.36\,{\rm GeV}^2$. However, at
the $1.5\sigma$ level this upper bound already becomes positive, which is not
very interesting since $\lambda_1$ is expected to be negative. $R_\tau$ is also
interesting for constraining new physics, e.g., possible (tree level) charged
Higgs contributions in the large $\tan\beta$ region~\cite{yuval}. 

\subsection{$B \to X_s \gamma$}

The mean photon energy over the region $E_0 < E_\gamma < E_\gamma^{\rm max}$ in
inclusive $B\to X_s\gamma$ decay for sufficiently small $E_0$  is sensitive to
$\bar\Lambda$ and independent of $\lambda_{1,2}$ when the decay rate is
expressed in terms of $m_B$ and $\bar\Lambda$ instead of $m_b$~\cite{AZ}.  We
expect that stringent constraints can be obtained on the nonperturbative
corrections to $m_\Upsilon$ when a measurement of the photon energy spectrum is
available with $E_0$ lowered to around $2\,$GeV~\cite{AZ,NaKa}.

Our method reduces the theoretical uncertainties in the total $B\to X_s\gamma$
decay rate only moderately. The $B\to X_s\gamma$ decay rate is usually
normalized to the semileptonic $B\to X_c e\bar\nu$ rate to eliminate the
$m_b^5$ factor from the prediction.  The two biggest ingredients of the total
theoretical error are the scale dependence of the $B\to X_c e\bar\nu$ rate
computation itself (without the $m_b^5$ factor), and to a lesser extent the
normalization to the $B\to X_c e\bar\nu$ rate due to the uncertainty of
$m_c/m_b$ and the experimental error.

\subsection{Nonleptonic $B$ decays}

The order $\lqcd^2/m_b^2$ corrections are taken from~\cite{incl,inclnl}.  We
neglect the strange quark mass, and the deviation of $|V_{cs}|^2 + |V_{cd}|^2$
and $|V_{ud}|^2 + |V_{us}|^2$ from unity.

The most interesting decay mode is $b\to c\bar c(s+d)$.  
\begin{eqnarray}\label{bccsd}
\Gamma(B\to X_{c\bar c(s+d)}) = {G_F^2 |V_{cb}|^2\over 64\pi^3}
  \bigg({m_\Upsilon\over2}\bigg)^5\, 0.212 \Big[ 1 &+& 
  0.156\epsilon + (0.074_{\rm BLM} + 0.157 L^2 + 0.046 L)\epsilon^2 \nonumber\\
&-& (0.45\lambda_2 + 0.41\lambda_1)/{\rm GeV}^2 \Big] \,, 
\end{eqnarray}
The variable $L=1$ is introduced to distinguish between terms that come from
different powers of $\ln(m_W/m_b)$. Equation~(\ref{bccsd}) has been written
with the renormalization group scaling factor expanded in powers of $\alpha_s
\ln(M_W/m_b)$. One can effectively include the complete leading logarithmic 
series by replacing the $0.157 L^2$ term by $0.113 L^2$. This replacement also
holds for Eqs.~(\ref{bcusd}), (\ref{buusd}) and (\ref{bucsd}). The
upsilon expansion does not affect the order $\epsilon^2 L$ and $\epsilon^2 L^2$
terms, i.e., they are the same as in the conventional $\alpha_s$ expansion. 
The behavior of the order $\epsilon$ and $(\epsilon^2)_{\rm BLM}$ terms have
improved significantly.  When written in terms of the pole mass, these terms
are $1 + 0.199\epsilon + 0.151_{\rm BLM}\epsilon^2$~\cite{LLSS}.  The
cancellation we find is rather nontrivial. The expansion of $m_b^5$ when
reexpressed in terms of $m_\Upsilon$ generates  positive contributions at
order $\epsilon^{1,2}$. This is why there was a significant cancellation in the
perturbation series for $b \to u e \bar \nu$ decay, since the conventional
perturbation series expansion for this decay had negative $\epsilon^{1,2}$
terms. On the  other hand, the conventional perturbation series for nonleptonic
$b \to c\bar c(s+d)$ has positive $\epsilon^{1,2}$ terms, and reexpressing
the overall $m_b^5$ factor in terms of $m_\Upsilon$ only makes them larger. It is the
expansions of the tree level $b\to c\bar c(s+d)$ phase space, which generates
large negative contributions at order $\epsilon^{1,2}$ to make the expansion
reasonably well-behaved.

For nonleptonic $b \to c \bar u (s+d)$ decay, the rate is
\begin{eqnarray}\label{bcusd}
\Gamma(B\to X_{c\bar u(s+d)}) = {G_F^2 |V_{cb}|^2\over 64\pi^3}
  \bigg({m_\Upsilon\over2}\bigg)^5\, 0.533 \Big[ 1 &-& 
  0.026\epsilon + (-0.006_{\rm BLM} + 0.157 L^2 + 0.099 L)\epsilon^2 \nonumber\\
&-& (0.28\lambda_2 + 0.12\lambda_1)/{\rm GeV}^2 \Big] \,, 
\end{eqnarray}
When written in terms of the pole mass, the order $\epsilon$ and
$(\epsilon^2)_{\rm BLM}$ terms are $1 - 0.048\epsilon - 0.045_{\rm
BLM}\epsilon^2$.

The decays mediated by $b\to u$ transitions are much less important, and
therefore we keep fewer terms than before.
\begin{equation}\label{buusd}
\Gamma(B\to X_{u\bar u(s+d)}) = {G_F^2 |V_{ub}|^2\over 64\pi^3}
  \bigg({m_\Upsilon\over2}\bigg)^5\, \Big[ 1 -
  0.045\epsilon + 0.157 L^2 \epsilon^2 \Big] \,, 
\end{equation}
For comparison, the order $\epsilon$ correction written in terms of the pole
mass is $1 - 0.099\epsilon$.

\begin{equation}\label{bucsd}
\Gamma(B\to X_{u\bar c(s+d)}) = {G_F^2 |V_{ub}|^2\over 64\pi^3}
  \bigg({m_\Upsilon\over2}\bigg)^5\, 0.533 \Big[ 1 +
  0.11\epsilon + 0.157 L^2 \epsilon^2 \Big] \,, 
\end{equation}
This is the only case where the upsilon expansion increases the order
$\epsilon$ correction compared to when it is written in terms of the pole mass,
$1 + 0.09\epsilon$.

\subsection{Semileptonic branching ratio and charm counting}

The theoretical prediction for the semileptonic $B$ branching ratio ${\cal
B}_{\rm SL}$ was discussed in~\cite{SLBR}.  It has been emphasized repeatedly
that the calculation of the $b\to c\bar c(s+d)$ rate is the least reliable. 
With our expansion in Eq.~(\ref{bccsd}), the ratio of the $(\epsilon^2)_{\rm
BLM}$ correction to the order $\epsilon$ term is significantly smaller than
that calculated in terms of the pole masses, so the $b\to c\bar c(s+d)$ rate
should be under better control. With our central values for the inclusive decay
rates given in the previous section we obtain
\begin{equation}\label{central1}
{\cal B}_{\rm SL} = 11.8\% \,, \qquad 
  n_c = 1.22 \,,
\end{equation}
for the semileptonic branching ratio and the average charm multiplicty, which
is consistent with other recent analyses. Here we have used
$|V_{ub}/V_{cb}|=0.1$.  The sensitivity of these results to $\lambda_1$, for
example, is small when it is varied in the range obtained in
Eq.~(\ref{lambda1}). If we had not  expanded the series of leading logarithms,
$n_c$ would remain unchanged, but ${\cal B}_{\rm SL}$ would increase to
$12.1\%$. Including a $+100$~MeV nonperturbative contribution to the $\Upsilon$
mass changes these values by a negligible amount.

\section{Conclusions}

We have shown that inclusive and exclusive $B$ decay rates can be predicted in
terms of the $\Upsilon(1S)$ mass instead of the $b$ quark mass.  It is crucial
to our analysis to use the upsilon expansion in $\epsilon$ rather than the
conventional expansion in powers of $\alpha_s$.  Our formulae relate only
physical quantities to one another. They result in smaller theoretical
uncertainties than existing numerical predictions, and the behavior of the
perturbation series is improved.  Moreover, the uncertainties can be estimated
without resorting to cumbersome arguments, and they can be checked using the
experimental data.  

In a previous paper~\cite{prl}, we had shown how our method works for inclusive
semileptonic $B$ decays. In this article, we have shown how it applies to
exclusive decays and to nonleptonic decays. In the new cases studied, the
upsilon expansion also improves the behavior of the perturbation series.  The
biggest theoretical uncertainty at present is a reliable bound on the effect 
of nonperturbative contributions to the $\Upsilon$ mass (other than the
linearly infrared sensitive piece of the pole mass). The best way of
determining this seems to be directly from the experimental data. The
comparison of $V_{cb}$ extracted using the upsilon expansion with previous
determinations already suggests that these contributions are not large. We are
investigating whether other processes can be used to provide an independent
test of this result.

\acknowledgements 

We thank Mike Luke and especially Mark Wise for useful discussions, and for
trying to keep us honest. This work was supported in part by the Department of
Energy under grant DOE-FG03-97ER40546, and by the National Science Foundation
under grant PHY-9457911. Fermilab is operated by Universities Research 
Association, Inc., under DOE contract  DE-AC02-76CH03000.

\end{document}